\documentclass[journal]{IEEEtran}
\IEEEoverridecommandlockouts
\usepackage{lineno}
\usepackage{cite}
\usepackage{hyperref}
\usepackage{amsmath,amssymb,amsfonts}
\usepackage{amsmath}
\usepackage{amsthm}

\usepackage{graphicx}
\usepackage{textcomp}
\usepackage{xcolor}
\usepackage{graphicx}
\usepackage{float}
\usepackage{subfigure}
\usepackage{amsmath}
\usepackage{amsfonts,amssymb}
\usepackage{mathrsfs}
\usepackage{mathtools}
\usepackage{algorithm}
\usepackage{algorithmicx}
\usepackage{algpseudocode}
\usepackage{bm}
\usepackage{multirow}
\usepackage{array}
\usepackage{amssymb}
\usepackage{amsmath}
\usepackage{cite}
\usepackage{url}
\usepackage{xcolor}
\usepackage{cite,graphicx,amsmath,amssymb}
\usepackage{subfigure}
\usepackage{fancyhdr}
\usepackage{mdwmath}
\usepackage{mdwtab}
\usepackage{caption}
\usepackage{amsthm}
\usepackage{setspace}
\usepackage{bm}
\usepackage{algorithm}
\usepackage{algpseudocode}
\usepackage{mathtools}
\usepackage{dsfont}
\usepackage{bbm}
\usepackage{yhmath}
\newtheorem{remark}{Remark}
\theoremstyle{definition}
\newtheorem{theorem}{Theorem}

\newtheorem{lemma}{Lemma}

\newtheorem{corollary}{Corollary}

\makeatletter
\newcommand{\biggg}{\bBigg@{3}}
\newcommand{\Biggg}{\bBigg@{3.5}}
\makeatother
\begin{document}
\title{Revealing the Impact of SIC in NOMA-ISAC}
\author{Chongjun~Ouyang, Yuanwei~Liu, and Hongwen~Yang
\thanks{C. Ouyang and H. Yang are with the School of Information and Communication Engineering, Beijing University of Posts and Telecommunications, Beijing, 100876, China (e-mail: \{DragonAim,yanghong\}@bupt.edu.cn).}
\thanks{Y. Liu is with the School of Electronic Engineering and Computer Science, Queen Mary University of London, London, E1 4NS, U.K. (e-mail: yuanwei.liu@qmul.ac.uk).}
}
\maketitle

\begin{abstract}
The impact of successive interference cancellation (SIC) in non-orthogonal multiple access integrated sensing and communications (NOMA-ISAC) is analyzed. A two-stage SIC-based framework is proposed to deal with the inter-communication user and inter-functionality interferences. The performance of sensing and communications (S\&C) is analyzed for two SIC orders, i.e., the communications-centric SIC and the sensing-centric SIC. For each design, diversity orders, high signal-to-noise ratio (SNR) slopes, and high-SNR power offsets of the sensing rate (SR) and communication rate (CR) are derived as insights. Analytical results indicate that \romannumeral1) the main influence of SIC order on the SR and CR lies in the high-SNR power offsets; \romannumeral2) ISAC provides more degrees of freedom than frequency-division S\&C (FDSAC). Numerical results show that the SR-CR region of ISAC entirely covers that of FDSAC.
\end{abstract}


\section{Introduction}
Integrated sensing and communications (ISAC) is a promising technology that allows sensing and communications (S\&C) to share the same spectrum and infrastructure \cite{Liu2022_JSAC}. This technology is believed to be more spectral- and hardware-efficient than existing frequency-division S\&C (FDSAC) techniques, in which S\&C functionalities exploit isolated frequency-hardware resources. Because of these advantages, ISAC is being paid lots of research attention \cite{Liu2022_JSAC}.

Recently, the performance of uplink ISAC systems has received growing attention; see \cite{Dong2022,Liu2022,Chiriyath2016,Zhang2022,Ouyang2022_CL,Ouyang2022_WCL} and the references therein. In this system, the dual-functional S\&C (DFSAC) receiver observes the sensing echo signal reflected by the targets and the digital signal sent by the communication users (CUs) simultaneously. We refer to this system as non-orthogonal multiple access (NOMA)-ISAC to highlight the double non-orthogonality therein, i.e., the non-orthogonality among the CUs and the non-orthogonality between the S\&C functionalities. The methods to deal with inter-CU interference (ICI) have been widely studied. As for inter-functionality interference (IFI), one trivial way to tackle it is to perform the sensing task and decode communication information in parallel by treating each other as a source of interference \cite{Dong2022,Liu2022}. This might limit the overall system performance \cite{Chiriyath2016}. As an improvement, the authors in \cite{Chiriyath2016,Zhang2022,Ouyang2022_CL,Ouyang2022_WCL} exploited a successive interference cancellation (SIC)-based framework to deal with the IFI. More specifically, the noisy communication signal is first decoded by treating the sensing signal as interference and then removed, leaving behind a communication interference-free sensing return.

It is worth noting that the SIC order adopted by the previous works is more favorable for sensing. Besides this SIC order, one could also first remove the sensing signal, and this SIC order will be more beneficial to communications. In a nutshell, the SIC order significantly influences the S\&C performance, which has yet to be fully understood. Motivated by this, we analyze the performance of NOMA-ISAC and discuss the influence of SIC ordering. The main contributions of this letter are listed as follows: \romannumeral1) We propose a two-stage SIC-based framework to tackle the ICI and IFI and analyze the outage probability (OP), ergodic communication rate (CR), and sensing rate (SR) achieved by two SIC orders, i.e., sensing-centric SIC (S-SIC) and communications-centric SIC (C-SIC); \romannumeral2) We perform asymptotic analyses in the high signal-to-noise ratio (SNR) regime to derive the diversity orders, high-SNR slopes, and high-SNR power offsets; \romannumeral3) We exploit the derived results to show that SIC ordering influences the CR and SR via shaping the high-SNR power offsets; \romannumeral4) We unveil that ISAC provides more degrees of freedom than FDSAC, and the SR-CR region of FDSAC is fully included in that of ISAC.

\section{System Model}
\subsection{Uplink NOMA-ISAC}
\begin{figure}[!t]
\centering
\setlength{\abovecaptionskip}{0pt}
\includegraphics[height=0.16\textwidth]{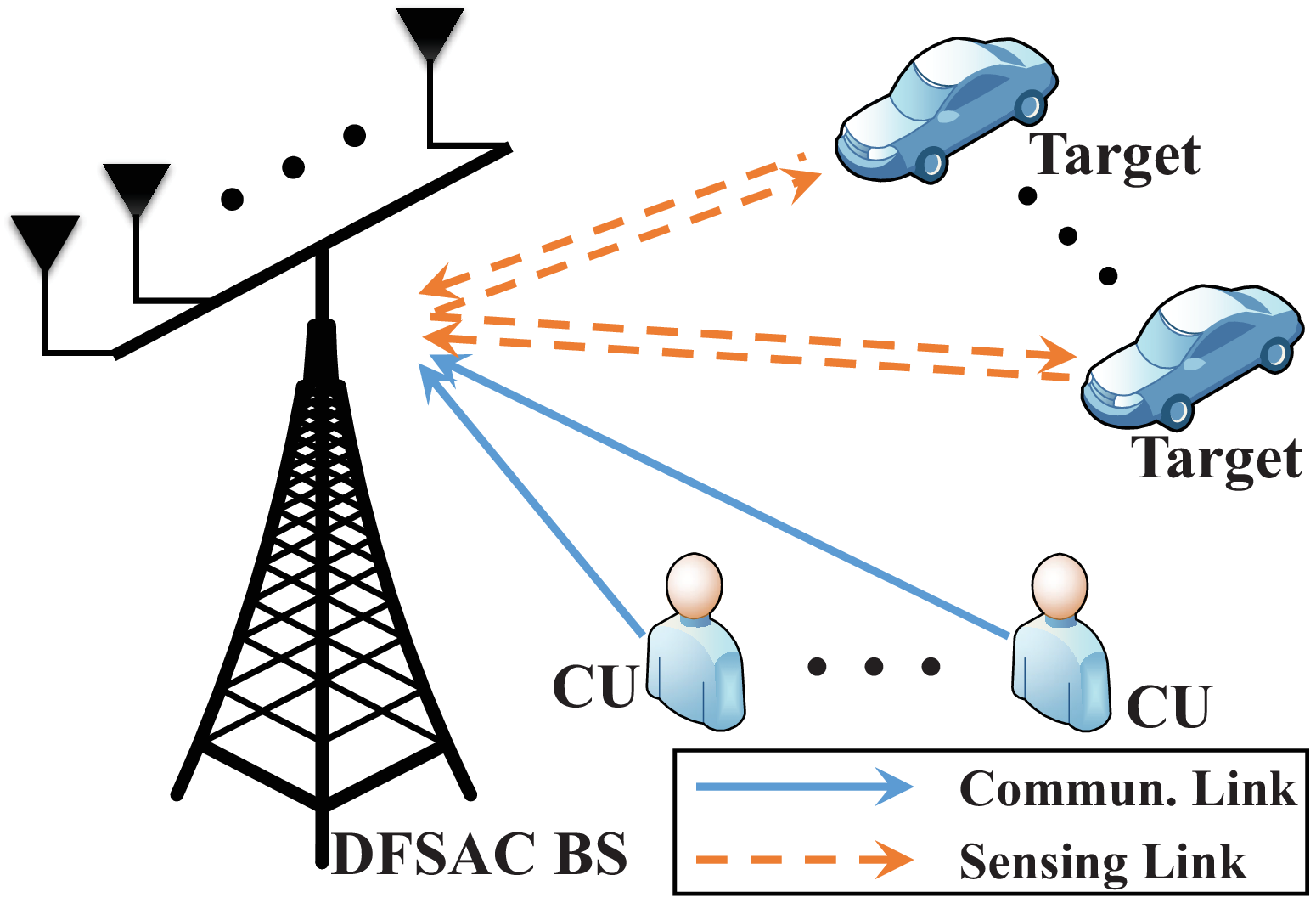}
\caption{Illustration of an uplink NOMA-ISAC system}
\label{System_Model}
\vspace{-20pt}
\end{figure}

A DFSAC base station (BS) serves $K$ uplink single-antenna CUs while simultaneously sensing the targets, as depicted in {\figurename} {\ref{System_Model}}. The uplink ISAC consists of two stages. Firstly, the BS broadcasts a predesigned sensing signal to the nearby environment. Secondly, the BS receives the sensing echoes reflected from the targets and the communication signals sent by the CUs. The sensing signal's time interval may be longer than the round-trip time of the sensing signal traveling between the BS and the targets; therefore, the two stages mentioned above could be coupled in time \cite{Chiriyath2016,Zhang2022,Ouyang2022_CL,Ouyang2022_WCL}.

The above arguments imply that the DFSAC BS should work in the full duplex mode. For brevity, we consider that the resultant self-interference is completely eliminated by equipping the BS with two sets of spatially well-separated antennas, i.e., $N$ transmit antennas and $M$ ($M\geq N$) receive antennas. We consider that $M\geq K$ and the CUs communicate with the BS under an uplink NOMA protocol.

The sensing and communication signals are assumed to be perfectly synchronized at the BS by using properly designed synchronization sequences. Consequently, the BS observes the following superposed S\&C signal:
{\setlength\abovedisplayskip{2pt}
\setlength\belowdisplayskip{2pt}
\begin{align}\label{Uplink_Basic_Model}
{\mathbf{Y}}=\sum\nolimits_{k=1}^{K}{{\mathbf{h}}}_k{\mathbf{x}}_k^{\mathsf{H}}+
{\mathbf{G}}^{\mathsf{H}}{\mathbf{S}}+{\mathbf{N}}^{\mathsf{H}}.
\end{align}
}The terms appearing in \eqref{Uplink_Basic_Model} are defined as follows:
\begin{itemize}
  \item ${\mathbf{S}}=\left[{\mathbf{s}}_1 \ldots {\mathbf{s}}_L\right]\in{\mathbbmss{C}}^{N\times L}$ ($L\geq M$, $L\geq N$) denotes the sensing waveform with $L$ being the pulse length and ${\mathbf{s}}_l\in{\mathbbmss C}^{N\times1}$ being the waveform at the $l$th time slot.
  \item $\mathbf{G}=[\mathbf{g}_{1} \ldots \mathbf{g}_M]\in{\mathbbmss{C}}^{N\times M}$ is the target response matrix with $\mathbf{g}_{m}\in{\mathbbmss{C}}^{N\times 1}$ representing the target response from the transmit array to the $m$th receive antenna.
  \item ${\mathbf{h}}_k=[h_{k,1},\ldots,h_{k,M}]^{\mathsf T}\in{\mathbbmss{C}}^{M\times1}$ is the channel vector from CU $k\in{\mathcal K}\triangleq\{1,\ldots,K\}$ to BS's receive antenna array, which is assumed to be known by the BS.
  \item ${\mathbf{x}}_k=\left[x_{k,1},\ldots,x_{k,L}\right]^{\mathsf{H}}\in{\mathbbmss{C}}^{L\times1}$ denotes the message sent by CU $k\in{\mathcal K}$ subject to ${\mathbbmss{E}}\{|x_{k,l}|^2\}= p_{\rm{c}}$ with $p_{\rm{c}}$ denoting the communication power.
  \item ${\mathbf{N}}=\left[{\mathbf{n}}_1 \ldots {\mathbf{n}}_L\right]^{\mathsf H}\in{\mathbbmss{C}}^{L\times M}$ is the Gaussian noise matrix containing $LM$ independent and identically distributed (i.i.d.) standard complex Gaussian elements (CGEs).
\end{itemize}

The target response matrix is modeled as \cite{Ouyang2022_CL,Ouyang2022_WCL}
{\setlength\abovedisplayskip{2pt}
\setlength\belowdisplayskip{2pt}
\begin{align}
\mathbf{G}=\sum\nolimits_{q}\beta_q{\mathbf{a}}(\theta_q){\mathbf{b}}^{\mathsf{H}}(\theta_q),
\end{align}
}where $\beta_q\sim{\mathcal{CN}}(0,\sigma_q^2)$ is the complex amplitude of the $q$th target with $\sigma_q^2$ representing the average strength, ${\mathbf{a}}(\theta_q)\in{\mathbbmss{C}}^{N\times1}$ and ${\mathbf{b}}(\theta_q)\in{\mathbbmss{C}}^{M\times1}$ are the associated transmit and receive array steering vectors, respectively, and $\theta_q$ is its direction of arrival. It is worth noting that the target response matrix $\mathbf G$ contains all the information about the targets, such as their location and number. Thus, target sensing can be equivalently treated as the estimation of the target response. Assuming that the receive antennas at the BS are widely separated, we have $\mathbf{g}_m\sim{\mathcal{CN}}({\mathbf{0}},\mathbf{R})$ for $m\in\{1,\ldots, M\}$ and ${\mathbbmss{E}}\{{\mathbf{g}}_m{\mathbf{g}}_{m'}^{\mathsf{H}}\}={\mathbf{0}}$ for $m\neq m'$. We also assume that the BS knows the correlation matrix ${\mathbf{R}}\in{\mathbbmss{C}}^{N\times N}$, which could be estimated based on multiple previous estimates. The communication channel is characterized by the Rayleigh fading model, which yields $\mathbf{h}_k\sim{\mathcal{CN}}({\mathbf{0}},\alpha_k\mathbf{I})$ for $k\in{\mathcal{K}}$ and ${\mathbbmss{E}}\{{\mathbf{h}}_k{\mathbf{h}}_{k'}^{\mathsf{H}}\}={\mathbf{0}}$ for $k\neq k'$, where $\alpha_k>0$ models the influence of large-scale path loss.

\subsection{A Two-Stage SIC-Based Framework}
\begin{figure}[!t]
\centering
\setlength{\abovecaptionskip}{0pt}
\includegraphics[height=0.23\textwidth]{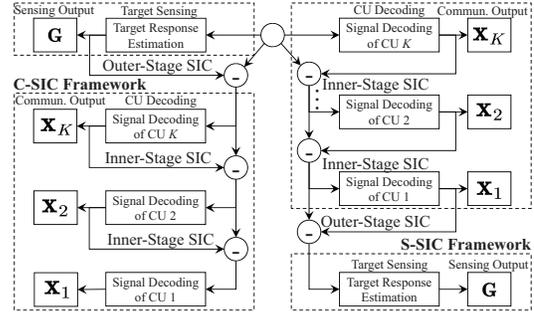}
\caption{Block diagram of the SIC-based framework}
\label{SIC_Model}
\vspace{-20pt}
\end{figure}

After receiving the superposed signal matrix ${\mathbf{Y}}$, the BS aims to decode the data information contained in the communication signal, $\left\{{\mathbf{x}}_k\right\}_{k=1}^{K}$, as well as extracting the environmental information contained in the target response, $\mathbf{G}$. To deal with the ICI and IFI, we propose a two-stage SIC-based framework, in which the inner-stage SIC deals with the ICI while the outer-stage SIC deals with the IFI, as detailed in {\figurename} {\ref{SIC_Model}}.

Without loss of generality, we assume that the CUs are arranged in an ascending order, i.e., $\lVert{\mathbf h}_K\rVert\geq\cdots\geq\lVert{\mathbf h}_1\rVert$. As {\figurename} {\ref{SIC_Model}} shows, the CUs with better channel conditions are decoded earlier in the inner-stage SIC. As for the outer-stage SIC, we consider two SIC orders. Under the first SIC order, the BS first senses the target response $\mathbf{G}$ by treating the communication signal as interference, and then ${\mathbf{G}}^{\mathsf{H}}{\mathbf{S}}$ is subtracted from $\mathbf{Y}$ with the rest part used for detecting the communication signal. Under the second SIC order, the BS first detects the communication signal $\left\{{\mathbf{x}}_k\right\}_{k=1}^{K}$ by treating the sensing signal as interference, and then $\sum\nolimits_{k=1}^{K}{{\mathbf{h}}}_k{\mathbf{x}}_k^{\mathsf{H}}$ is subtracted from $\mathbf{Y}$ with the rest part used for sensing the target response. Clearly, the first SIC order yields a better communication performance while the second one yields a better sensing performance. Motivated by this, we refer to these two SIC orders as the C-SIC and S-SIC, respectively.

Having established the fundamental model of the two-stage SIC-based framework, we now move to the discussion on the S\&C performance achieved by these two SIC orders.

\section{Performance of NOMA-ISAC}
\subsection{ISAC With Communications-Centric SIC}
\subsubsection{Performance of Sensing}
Let us first study the sensing performance achieved by the C-SIC. As mentioned earlier, the objective of sensing is extracting the environmental information contained in the target response $\mathbf{G}$ by observing $\mathbf{Y}$. Since the sensing signal $\mathbf{S}$ is designed in advance and known by the BS, we can use the mutual information (MI) between $\mathbf{Y}$ and $\mathbf{G}$ conditioned on $\mathbf{S}$ to evaluate how much environmental information can be extracted. This MI is also referred to as the sensing MI, which characterizes the information-theoretic limits of sensing. On this basis, from an information-theoretic perspective, we exploit the SR to evaluate the sensing performance, which is defined as the sensing MI per unit time \cite{Ouyang2022_CL,Ouyang2022_WCL}. Assuming that each sensing waveform symbol lasts $1$ unit time, we write the SR as $L^{-1}I(\mathbf{Y};\mathbf{G}|\mathbf{S})$, where $I\left(X;Y|Z\right)$ denotes the MI between $X$ and $Y$ conditioned on $Z$.

From a worst-case design perspective, the aggregate interference-plus-noise ${\mathbf Z}=\sum_{k=1}^{K}{{\mathbf{h}}}_k{\mathbf{x}}_k^{\mathsf{H}}+{\mathbf{N}}^{\mathsf{H}}$ is treated as the Gaussian noise \cite{Hassibi2003}. We further consider that the sensing signal is designed by exploiting the statistical information of the communication channels and that the communication symbols sent at different time slots are statistically uncorrelated, i.e., ${\mathbbmss E}\{x_{k,l}x_{k,l'}^{*}\}=0$ for $l\neq l'$. Under this consideration, we conclude the following lemma.
\vspace{-5pt}
\begin{lemma}\label{Sensing_MI_C_SIC_Lemma}
The SR achieved by the C-SIC is given by $\frac{M}{L}\log_2\det(\mathbf{I}+\sigma_{\rm{c}}^{-2}{\mathbf{S}}^{\mathsf{H}}{\mathbf{R}}
{\mathbf{S}})$ with $\sigma_{\rm{c}}^2=1+{p_{\rm{c}}}\sum_{k=1}^{K}\alpha_k$.
\end{lemma}
\vspace{-5pt}
\begin{IEEEproof}
Please refer to Appendix \ref{Proof_Sensing_MI_C_SIC_Lemma} for more details.
\end{IEEEproof}
It is worth noting that the SR is a function of the sensing signal $\mathbf{S}$ and the maximum SR can be expressed as
{\setlength\abovedisplayskip{2pt}
\setlength\belowdisplayskip{2pt}
\begin{align}\label{SR_Basic_C_SIC}
\mathcal{R}_{\rm{s}}^{\rm{c}}=\max\nolimits_{\mathsf{tr}\left({\mathbf{S}}{\mathbf{S}}^{\mathsf{H}}\right)\leq Lp_{\rm{s}}}{M}{L}^{-1}\log_2\det(\mathbf{I}+\sigma_{\rm{c}}^{-2}{\mathbf{S}}^{\mathsf{H}}{\mathbf{R}}
{\mathbf{S}}),
\end{align}
}where $p_{\rm{s}}$ is the per-symbol power budget of the sensing signal. For analytical tractability, we assume that ${\mathbf{R}}\succ{\mathbf{0}}$. The following theorem provides an exact expression for $\mathcal{R}_{\rm{s}}^{\rm{c}}$ as well as its high-SNR approximation.
\vspace{-5pt}
\begin{theorem}\label{Sensing_Rate_C_SIC_Theorem}
The maximum SR achieved by the C-SIC is
{\setlength\abovedisplayskip{2pt}
\setlength\belowdisplayskip{2pt}
\begin{align}
\mathcal{R}_{\rm{s}}^{\rm{c}}={M}{L}^{-1}\sum\nolimits_{n=1}^{N}\log_2\left(1+{\sigma_{\rm{c}}^{-2}}\lambda_ns_n^{\star}\right),
\end{align}
}where $\{\lambda_n>0\}_{n=1}^{N}$ are the eigenvalues of matrix ${\mathbf{R}}$ and $s_{n}^{\star}=\max\{0,{\nu}^{-1}-{\sigma_{\rm{c}}^2}{\lambda_n^{-1}}\}$ with $\sum_{n=1}^{N}\max\{0,{\nu}^{-1}-{\sigma_{\rm{c}}^2}{\lambda_n^{-1}}\}=Lp_{\rm{s}}$. The maximum SR is attained when ${\mathbf{S}}{\mathbf{S}}^{\mathsf{H}}={\mathbf{U}}^{\mathsf{H}}{\bm\Delta}_{\rm{c}}^{\star}{\mathbf{U}}$, where ${\mathbf{U}}^{\mathsf{H}}{\mathsf{diag}}\left\{\lambda_1,\cdots,\lambda_{N}\right\}{\mathbf{U}}$ denotes the eigendecomposition (ED) of ${\mathbf{R}}$ and ${\bm\Delta}_{\rm{c}}^{\star}={\mathsf{diag}}\left\{s_1^{\star},\cdots,s_N^{\star}\right\}$.
When $p_{\rm{s}}\rightarrow\infty$, the maximum achievable SR satisfies
{\setlength\abovedisplayskip{2pt}
\setlength\belowdisplayskip{2pt}
\begin{align}\label{Sensing_Rate_C_SIC_Asymptotic}
\mathcal{R}_{\rm{s}}^{\rm{c}}
\approx{\mathcal S}_{\rm{s}}^{\rm{c}}\left(\log_2{p_{\rm{s}}}-{\mathcal L}_{\rm{s}}^{\rm{c}}\right),
\end{align}
}where ${\mathcal S}_{\rm{s}}^{\rm{c}}=\frac{NM}{L}$ and ${\mathcal L}_{\rm{s}}^{\rm{c}}=\frac{1}{N}\sum\nolimits_{n=1}^{N}\log_2\left(\frac{N\sigma_{\rm{c}}^2}{L\lambda_n}\right)$.
\end{theorem}
\vspace{-5pt}
\begin{IEEEproof}
Similar to the proof of \cite[Theorem 3]{Ouyang2022_WCL}.
\end{IEEEproof}
\vspace{-5pt}
\begin{remark}
The results in \eqref{Sensing_Rate_C_SIC_Asymptotic} suggest that the high-SNR slope and the high-SNR power offset of the SR achieved by the C-SIC are given by $\frac{NM}{L}$ and $\frac{1}{N}\sum\nolimits_{n=1}^{N}\log_2\left(\frac{N\sigma_{\rm{c}}^2}{L\lambda_n}\right)$, respectively.
\end{remark}
\vspace{-5pt}
\subsubsection{Performance of Communications}
After the target response is sensed, the sensing echo signal ${\mathbf{G}}^{\mathsf{H}}{\mathbf{S}}$ can be subtracted from the superposed S\&C signal $\mathbf Y$ shown in \eqref{Uplink_Basic_Model}. Then, the rest part will be used to recover the data information contained in $\left\{{\mathbf{x}}_k\right\}_{k=1}^{K}$. In order to unveil the performance upper bound of our considered uplink NOMA-ISAC, we assume that the sensing echo signal is perfectly removed via the C-SIC. On this basis, the BS observes the following signal after the C-SIC:
{\setlength\abovedisplayskip{2pt}
\setlength\belowdisplayskip{2pt}
\begin{align}
{\mathbf{Y}}_{\rm{c}}=\sum\nolimits_{k=1}^{K}{{\mathbf{h}}}_k{\mathbf{x}}_k^{\mathsf{H}}+{\mathbf{N}}^{\mathsf{H}}.
\end{align}
}Afterwards, the inner-stage minimum mean-square error (MMSE)-SIC-based decoder is utilized to detect the $K$ uplink NOMA signals $\{{\mathbf{x}}_k\}_{k=1}^{K}$, as shown in {\figurename} {\ref{SIC_Model}}. The signal-to-interference-plus-noise ratio (SINR) of CU $k$ is $\gamma_{k}^{\rm{c}}={\mathbf h}_k^{\mathsf H}({\mathbf I}+\sum_{i=1}^{k-1}p_{\rm c}{\mathbf h}_i{\mathbf h}_i^{\mathsf H})^{-1}{\mathbf h}_k$. By defining ${\mathbf{H}}=\left[{{\mathbf{h}}}_1 \ldots {{\mathbf{h}}}_K\right]\in{\mathbbmss{C}}^{M\times K}$, the sum-CR can be calculated as \cite{Heath2018}
{\setlength\abovedisplayskip{2pt}
\setlength\belowdisplayskip{2pt}
\begin{align}
\overline{\mathcal{R}}_{\rm{c}}^{\rm{c}}=\sum\nolimits_{k=1}^{K}\log_2(1+\gamma_{k}^{\rm{c}})
=\log_2\det(\mathbf{I}+p_{\rm{c}}{{\mathbf{H}}}{{\mathbf{H}}}^{\mathsf{H}}).
\end{align}
}Note that the MMSE-SIC decoder is sum-CR capacity-achieving. Besides, the sum-CR is always the same, no matter which decoding order is used \cite{Heath2018}. We next evaluate the communication performance by the OP and ergodic CR (ECR).

The OP is defined as the probability of the sum-CR being lower than a target rate $\mathcal{R}$, which can be written as
{\setlength\abovedisplayskip{2pt}
\setlength\belowdisplayskip{2pt}
\begin{align}
\mathcal{P}_{\rm{c}}^{\rm{c}}=\Pr(\overline{\mathcal{R}}_{\rm{c}}^{\rm{c}}<\mathcal{R})=
\Pr(\det(\mathbf{I}+p_{\rm{c}}{{\mathbf{H}}}{{\mathbf{H}}}^{\mathsf{H}})<2^{\mathcal{R}}).
\end{align}
}Yet, it is challenging to derive any tractable closed-form expressions of the OP $\mathcal{P}_{\rm{c}}^{\rm{c}}$. To glean further insights, we characterize the OP in the high-SNR regime as follows.
\vspace{-5pt}
\begin{theorem}\label{OP_C_SIC_Theorem}
As $p_{\rm{c}}\rightarrow\infty$, the OP satisfies $\mathcal{P}_{\rm{c}}^{\rm{c}}\simeq{\mathcal{O}}(p_{\rm{c}}^{-MK})$.
\end{theorem}
\vspace{-5pt}
\begin{IEEEproof}
Please refer to \cite{Heath2018} for more details.
\end{IEEEproof}
\vspace{-5pt}
\begin{remark}
The results in Theorem \ref{OP_C_SIC_Theorem} suggest the diversity order of the sum-CR achieved by the C-SIC is $KM$.
\end{remark}
\vspace{-5pt}
The sum ECR is defined as ${\mathcal{R}}_{{\rm{c}}}^{\rm{c}}={\mathbbmss{E}}\{\overline{\mathcal{R}}_{\rm{c}}^{\rm{c}}\}$. Notice that deriving a tractable expression for ${\mathcal{R}}_{{\rm{c}}}^{\rm{c}}$ is also a challenging task. We hence consider high-SNR limit of the sum ECR.
\vspace{-5pt}
\begin{theorem}\label{Theorem_ECR_C_SIC_Asymptotic}
When $p_{\text{c}}\rightarrow\infty$, the sum ECR satisfies
{\setlength\abovedisplayskip{2pt}
\setlength\belowdisplayskip{2pt}
\begin{align}\label{ECR_C_SIC_Asymptotic}
{\mathcal{R}}_{{\rm{c}}}^{\rm{c}}\approx {\mathcal S}_{\rm{c}}^{\rm{c}}\left(\log_2{p_{\rm{c}}}-{\mathcal L}_{\rm{c}}^{\rm{c}}\right),
\end{align}
}where ${\mathcal S}_{\rm{c}}^{\rm{c}}=K$, ${\mathcal L}_{\rm{c}}^{\rm{c}}=\frac{-1}{K}\sum\nolimits_{k=1}^{K}\left(\log_2{\alpha_k}+\sum\nolimits_{a=1}^{M-k}
\frac{a^{-1}-\emph{C}}{\ln{2}}\right)$, and $\emph{C}$ is the Euler constant.
\end{theorem}
\vspace{-5pt}
\begin{IEEEproof}
Please refer to Appendix \ref{Proof_Theorem_ECR_C_SIC_Asymptotic} for more details.
\end{IEEEproof}
\vspace{-5pt}
\begin{remark}
The results in \eqref{ECR_C_SIC_Asymptotic} suggest that the high-SNR slope and the high-SNR power offset of the sum ECR achieved by the C-SIC are given by $K$ and ${\mathcal L}_{\rm{c}}^{\rm{c}}$, respectively.
\end{remark}
\vspace{-5pt}
\subsection{ISAC With Sensing-Centric SIC}
Having investigated the S\&C performance achieved by the C-SIC, we now move to the S-SIC.
\subsubsection{Performance of Communications}
At the $l$th time slot, the BS observes the signal vector as follows:
{\setlength\abovedisplayskip{2pt}
\setlength\belowdisplayskip{2pt}
\begin{align}
{\mathbf{y}}_{l}=\sum\nolimits_{k=1}^{K}{\mathbf{h}}_{k}x_{k,l}+
{\mathbf{G}}^{\mathsf{H}}{\mathbf{s}}_l+{\mathbf{n}}_{l},
\end{align}
}where ${\mathbf{n}}_{l}\sim{\mathcal{CN}}\left({\mathbf{0}},{\mathbf{I}}\right)$. As {\figurename} {\ref{SIC_Model}} shows, we exploit the inner-stage MMSE-SIC decoder to detect the information bits, which is sum-CR capacity-achieving \cite{Heath2018}. From a worst-case design perspective, the aggregate interference-plus-noise term ${\mathbf{a}}_l={\mathbf{G}}^{\mathsf{H}}{\mathbf{s}}_l+{\mathbf{n}}_{l}$ is treated as the Gaussian noise \cite{Hassibi2003}. The uplink sum-CR at the $l$th time slot is thus given as follows.
\vspace{-5pt}
\begin{lemma}\label{Commun_MI_S_SIC_Lemma}
The sum-CR achieved by the S-SIC is given by $\overline{\mathcal{R}}_{{\rm{c}},l}^{\rm{s}}=\log_2\det({\mathbf{I}}+p_{\rm{c}}{\varrho_l^{-2}}{\mathbf{H}}{\mathbf{H}}^{\mathsf{H}})$ with $\varrho_l^2=1+|{\mathbf{s}}_l^{\mathsf{H}}{\mathbf{R}}{\mathbf{s}}_l|$.
\end{lemma}
\vspace{-5pt}
\begin{IEEEproof}
Please refer to Appendix \ref{Proof_Commun_MI_S_SIC_Lemma} for more details.
\end{IEEEproof}
The uplink sum-CR varies with the time slot index. We thus use the expectation of $\overline{\mathcal{R}}_{{\rm{c}},l}^{\rm{s}}$ in terms of $l$ to evaluate the communication performance, viz. $\overline{\mathcal{R}}_{{\rm{c}}}^{\rm{s}}=\frac{1}{L}\sum_{l=1}^{L}\overline{\mathcal{R}}_{{\rm{c}},l}^{\rm{s}}$.

The OP of the uplink sum-CR can be written as ${\mathcal{P}}_{{\rm{c}}}^{\rm{s}}=\Pr(\overline{\mathcal{R}}_{{\rm{c}}}^{\rm{s}}<{\mathcal{R}})$. Using the results in Theorem \ref{OP_C_SIC_Theorem} and the Sandwich theorem, we characterize the high-SNR behaviour of the OP as follows.
\vspace{-5pt}
\begin{theorem}\label{OP_S_SIC_Theorem}
As $p_{\rm{c}}\rightarrow\infty$, the OP satisfies $\mathcal{P}_{\rm{c}}^{\rm{s}}\simeq{\mathcal{O}}(p_{\rm{c}}^{-MK})$.
\end{theorem}
\vspace{-5pt}
\vspace{-5pt}
\begin{remark}
The results in Theorem \ref{OP_S_SIC_Theorem} suggest the diversity order of the sum-CR achieved by the S-SIC is $KM$.
\end{remark}
\vspace{-5pt}
The uplink ECR is given by ${\mathcal{R}}_{{\rm{c}}}^{\rm{s}}={\mathbbmss{E}}\{\overline{\mathcal{R}}_{{\rm{c}}}^{\rm{s}}\}$. Using similar steps as those outlined in Appendix \ref{Proof_Theorem_ECR_C_SIC_Asymptotic}, we get Theorem \ref{Theorem_ECR_S_SIC_Asymptotic}.
\vspace{-5pt}
\begin{theorem}\label{Theorem_ECR_S_SIC_Asymptotic}
When $p_{\text{c}}\rightarrow\infty$, the sum ECR satisfies
{\setlength\abovedisplayskip{2pt}
\setlength\belowdisplayskip{2pt}
\begin{align}\label{ECR_S_SIC_Asymptotic}
{\mathcal{R}}_{{\rm{c}}}^{\rm{s}}\approx {\mathcal S}_{\rm{c}}^{\rm{s}}\left(\log_2{p_{\rm{c}}}-{\mathcal L}_{\rm{c}}^{\rm{s}}\right),
\end{align}
}where ${\mathcal S}_{\rm{c}}^{\rm{s}}=K$ and ${\mathcal L}_{\rm{c}}^{\rm{s}}=\frac{1}{L}\sum\nolimits_{l=1}^{L}\log_2{{\varrho_l^2}}-\frac{1}{K}\sum\nolimits_{k=1}^{K}\left(\log_2{\alpha_k}+\sum\nolimits_{a=1}^{M-k}
\frac{a^{-1}-\emph{C}}{\ln{2}}\right)$.
\end{theorem}
\vspace{-5pt}
\vspace{-5pt}
\begin{remark}
The results in \eqref{ECR_S_SIC_Asymptotic} suggest that the high-SNR slope and the high-SNR power offset of the sum ECR achieved by the S-SIC are given by $K$ and ${\mathcal L}_{\rm{c}}^{\rm{s}}$, respectively.
\end{remark}
\vspace{-5pt}
\subsubsection{Performance of Sensing}
After decoding all the information bits sent by CUs, the BS can remove $\sum\nolimits_{k=1}^{K}{{\mathbf{h}}}_k{\mathbf{x}}_k^{\mathsf{H}}$ from ${\mathbf{Y}}$ by virtue of the S-SIC. Then, the rest part is used to sense the target response. To unveil the performance upper bound, we assume that $\sum\nolimits_{k=1}^{K}{{\mathbf{h}}}_k{\mathbf{x}}_k^{\mathsf{H}}$ is perfectly removed. On this basis, the BS observes the following signal
after the S-SIC:
{\setlength\abovedisplayskip{2pt}
\setlength\belowdisplayskip{2pt}
\begin{align}
{\mathbf{Y}}_{\rm{s}}={\mathbf{G}}^{\mathsf{H}}{\mathbf{S}}+{\mathbf{N}}^{\mathsf{H}}.
\end{align}
}The maximum SR is thus obtained by setting $p_{\rm c}$ in \eqref{SR_Basic_C_SIC} as $0$:
{\setlength\abovedisplayskip{2pt}
\setlength\belowdisplayskip{2pt}
\begin{align}
{\mathcal{R}}_{{\rm{s}}}^{\rm{s}}=\max\nolimits_{\mathsf{tr}\left({\mathbf{S}}{\mathbf{S}}^{\mathsf{H}}\right)\leq Lp_{\rm{s}}}{M}{L}^{-1}\log_2\det(\mathbf{I}+{\mathbf{S}}^{\mathsf{H}}{\mathbf{R}}
{\mathbf{S}}).
\end{align}
}By the method we derive Theorem \ref{Sensing_Rate_C_SIC_Theorem}, we obtain Theorem \ref{Theorem_SR_ER}.
\vspace{-5pt}
\begin{theorem}\label{Theorem_SR_ER}
The maximum SR achieved by the S-SIC is
{\setlength\abovedisplayskip{2pt}
\setlength\belowdisplayskip{2pt}
\begin{align}
{\mathcal{R}}_{{\rm{s}}}^{\rm{s}}={M}{L}^{-1}\sum\nolimits_{n=1}^{N}\log_2\left(1+\lambda_na_n^{\star}\right),
\end{align}
}where $a_{n}^{\star}=\max\{0,{\nu}^{-1}-{\lambda_n^{-1}}\}$ with $\sum_{n=1}^{N}\max\{0,{\nu}^{-1}-{\lambda_n^{-1}}\}=Lp_{\rm{s}}$. The SR is maximized when ${\mathbf{S}}{\mathbf{S}}^{\mathsf{H}}={\mathbf{U}}^{\mathsf{H}}{\bm\Delta}_{\rm s}^{\star}{\mathbf{U}}$ with ${\bm\Delta}_{\rm s}^{\star}={\mathsf{diag}}\{a_1^{\star},\cdots,a_N^{\star}\}$. When $p_{\rm{s}}\rightarrow\infty$, we have
{\setlength\abovedisplayskip{2pt}
\setlength\belowdisplayskip{2pt}
\begin{align}\label{Sensing_Rate_S_SIC_Asymptotic}
\mathcal{R}_{\rm{s}}^{\rm{s}}
\approx{\mathcal S}_{\rm{s}}^{\rm{s}}\left(\log_2{p_{\rm{s}}}-{\mathcal L}_{\rm{s}}^{\rm{s}}\right),
\end{align}
}where ${\mathcal S}_{\rm{s}}^{\rm{s}}=\frac{NM}{L}$ and ${\mathcal L}_{\rm{s}}^{\rm{s}}=\frac{1}{N}\sum\nolimits_{n=1}^{N}\log_2\left(\frac{N}{L\lambda_n}\right)$.
\end{theorem}
\vspace{-5pt}
\vspace{-5pt}
\begin{remark}
The results in \eqref{Sensing_Rate_S_SIC_Asymptotic} suggest that the high-SNR slope and the high-SNR power offset of the SR achieved by the S-SIC is given by $\frac{NM}{L}$ and $\frac{1}{N}\sum\nolimits_{n=1}^{N}\log_2\left(\frac{N}{L\lambda_n}\right)$, respectively.
\end{remark}
\vspace{-5pt}
\subsection{Discussion on SIC Ordering}\label{Discussion_SIC_Order}
In the following, we discuss the influence of SIC ordering. Let us first focus on the SR. By comparing \eqref{Sensing_Rate_C_SIC_Asymptotic} with \eqref{Sensing_Rate_S_SIC_Asymptotic}, we obtain the following results.
\vspace{-5pt}
\begin{remark}
The fact of ${\mathcal S}_{\rm{s}}^{\rm{c}}={\mathcal S}_{\rm{s}}^{\rm{s}}=\frac{NM}{L}$ suggests that the SIC ordering has no influence on the high-SNR slope of the SR.
\end{remark}
\vspace{-5pt}
\vspace{-5pt}
\begin{corollary}\label{Corollary_SR_Power_Compare}
As $p_{\rm s}\rightarrow\infty$, we can obtain
{\setlength\abovedisplayskip{2pt}
\setlength\belowdisplayskip{2pt}
\begin{align}
\mathcal{R}_{\rm{s}}^{\rm{s}}-\mathcal{R}_{\rm{s}}^{\rm{c}}&\approx
{NM}{L}^{-1}\left({\mathcal L}_{\rm{s}}^{\rm{c}}-{\mathcal L}_{\rm{s}}^{\rm{s}}\right)={NM}{L}^{-1}\log_2{{\sigma}_{\rm c}^2}\\
&=\frac{NM}{L}\log_2\left(1+{p_{\rm{c}}}\sum\nolimits_{k=1}^{K}\alpha_k\right)\triangleq{\mathcal E}_{\rm s}>0.
\end{align}
}\end{corollary}
\vspace{-5pt}
\vspace{-5pt}
\begin{remark}\label{Discussion_SIC_Sensing}
The above results suggest that the SIC ordering influences the SR via shaping its high-SNR power offset. More specifically, S-SIC is superior to C-SIC in terms of the achievable SR by an SNR gap of ${\mathcal E}_{\rm s}$ in 3-dB units \cite{Heath2018}. In other words, to achieve the same SR as S-SIC, C-SIC has to consume more SNR of ${\mathcal E}_{\rm s}$ in 3-dB units to resist the IFI \cite{Heath2018}.
\end{remark}
\vspace{-5pt}
Note that the gap ${\mathcal E}_{\rm s}$ is a monotone increasing function of the communication power budget $p_{\rm c}$, which is as expected.

We next consider the sum-CR. By comparing the results in Theorems \ref{OP_C_SIC_Theorem} and \ref{OP_S_SIC_Theorem}, we find that the SIC ordering does not influence the diversity order of the sum-CR. As is widely known, the high-SNR OP is determined by the array gain and diversity order \cite{Heath2018}. Since the SIC ordering does not influence the diversity order, we can infer that the SIC ordering could affect the array gain. Unfortunately, quantifying the gap with respect to the array gain is challenging, which is left as a potential direction for future work. In the sequel, we compare \eqref{ECR_C_SIC_Asymptotic} with \eqref{ECR_S_SIC_Asymptotic}, which leads to the following results.
\vspace{-5pt}
\begin{remark}
The fact of ${\mathcal S}_{\rm{c}}^{\rm{c}}={\mathcal S}_{\rm{c}}^{\rm{s}}=K$ suggests that the SIC ordering has no influence on the high-SNR slope of the ECR.
\end{remark}
\vspace{-5pt}
\vspace{-5pt}
\begin{corollary}\label{Corollary_CR_Power_Compare}
As $p_{\rm c}\rightarrow\infty$, we can obtain
{\setlength\abovedisplayskip{2pt}
\setlength\belowdisplayskip{2pt}
\begin{align}
\mathcal{R}_{\rm{c}}^{\rm{c}}-\mathcal{R}_{\rm{c}}^{\rm{s}}&\approx
K\left({\mathcal L}_{\rm{c}}^{\rm{s}}-{\mathcal L}_{\rm{c}}^{\rm{c}}\right)=KL^{-1}\sum\nolimits_{l=1}^{L}\log_2{{\varrho_l^2}}\\
&=K{L}^{-1}\sum\nolimits_{l=1}^{L}\!\log_2(1+|{\mathbf{s}}_l^{\mathsf{H}}{\mathbf{R}}{\mathbf{s}}_l|)\triangleq{\mathcal E}_{\rm c}>0.
\end{align}
}\end{corollary}
\vspace{-5pt}
\vspace{-5pt}
\begin{remark}\label{Discussion_SIC_Communications}
The above results suggest that the SIC ordering influences the sum ECR via shaping its high-SNR power offset. Particularly, C-SIC is superior to S-SIC in terms of the achievable ECR by an SNR gap of ${\mathcal E}_{\rm c}$ in 3-dB units \cite{Heath2018}. %Or in other words, to achieve the same ECR as C-SIC, S-SIC has to consume more SNR of ${\mathcal E}_{\rm c}$ in 3-dB units to resist the interference from the sensing echo signal. Intuitively, this SNR gap ${\mathcal E}_{\rm c}$ increases with the sensing power budget $p_{\rm s}$.
\end{remark}
\vspace{-5pt}
\subsubsection*{Summary}The above arguments imply that the SIC order affects the SR and CR by influencing the array gains and high-SNR power offsets.
\subsection{Rate Region Characterization}
After analyzing the S\&C performance and discussing the influence of SIC ordering, we next characterize the SR-CR region achieved by NOMA-ISAC.

Notice that $({\mathcal{R}}_{{\rm{s}}}^{\rm{s}},\overline{\mathcal{R}}_{{\rm{c}}}^{\rm{s}})$ and $({\mathcal{R}}_{{\rm{s}}}^{\rm{c}},\overline{\mathcal{R}}_{\rm{c}}^{\rm{c}})$ represent the maximum achievable rate tuples of S-SIC and C-SIC, respectively. Besides, ${\mathcal{R}}_{{\rm{s}}}^{\rm{s}}$ and $\overline{\mathcal{R}}_{{\rm{c}}}^{\rm{c}}$ represent the maximum SR and sum-CR achieved in NOMA-ISAC, respectively. By exploiting the celebrated time-sharing strategy, namely applying the S-SIC with probability $p$, while applying the C-SIC with probability $1-p$, we find that the rate tuple $({\mathcal{R}}_{{\rm{s}}}^{p},\overline{\mathcal{R}}_{{\rm{c}}}^{p})$ is attainable for $p\in[0,1]$, where ${\mathcal{R}}_{{\rm{s}}}^{p}=p{\mathcal{R}}_{{\rm{s}}}^{\rm{s}}+(1-p){\mathcal{R}}_{{\rm{s}}}^{\rm{c}}$ and $\overline{\mathcal{R}}_{{\rm{c}}}^{p}=p\overline{\mathcal{R}}_{{\rm{c}}}^{\rm{s}}+(1-p)\overline{\mathcal{R}}_{{\rm{c}}}^{\rm{c}}$. Let ${\mathcal{R}}^{\rm{s}}$ and ${\mathcal{R}}^{\rm{c}}$ denote the achievable SR and sum ECR, respectively. Then, the rate region achieved by ISAC reads
{\setlength\abovedisplayskip{2pt}
\setlength\belowdisplayskip{2pt}
\begin{align}
\mathcal{C}_{\rm{i}}=\left\{\left({\mathcal{R}}^{\rm{s}},{\mathcal{R}}^{\rm{c}}\right)|{\mathcal{R}}^{\rm{s}}\!\in\!\left[0,\mathcal{R}_{\rm{s}}^{p}\right],
{\mathcal{R}}^{\rm{c}}\!\in\!\left[0,\mathcal{R}_{\rm{c}}^{p}\right],p\!\in\!\left[0,\!1\right]\right\},\label{Rate_Regio_ISAC}
\end{align}
}where $\mathcal{R}_{\rm{c}}^{p}={\mathbbmss E}\{\overline{\mathcal{R}}_{\rm{c}}^{p}\}$. By the Sandwich theorem, we get the following results.
\vspace{-5pt}
\begin{corollary}
For a given $p\in[0,1]$, we have $\lim_{p_{\rm c}\rightarrow\infty}\Pr({\overline{\mathcal{R}}_{\rm{c}}^{p}}<\mathcal{R})\simeq{\mathcal{O}}(p_{\rm{c}}^{-MK})$.
\end{corollary}
\vspace{-5pt}
\vspace{-5pt}
\begin{corollary}
For a given $p\in[0,1]$, we have $\lim_{p_{\rm c}\rightarrow\infty}\mathcal{R}_{\rm{c}}^{p}\simeq K\left(\log_2{p_{\rm{c}}}-{\mathcal L}_{\rm{c}}^{p}\right)$ with ${\mathcal L}_{\rm{c}}^{p}=(1-p){\mathcal L}_{\rm{c}}^{\rm{c}}+p{\mathcal L}_{\rm{c}}^{\rm{s}}\in[{\mathcal L}_{\rm{c}}^{\rm{c}},{\mathcal L}_{\rm{c}}^{\rm{s}}]$.
\end{corollary}
\vspace{-5pt}
\vspace{-5pt}
\begin{corollary}
For a given $p\in[0,1]$, we have $\lim_{p_{\rm s}\rightarrow\infty}\mathcal{R}_{\rm{s}}^{p}\simeq \frac{NM}{L}\left(\log_2{p_{\rm{s}}}-{\mathcal L}_{\rm{s}}^{p}\right)$ with ${\mathcal L}_{\rm{s}}^{p}=p{\mathcal L}_{\rm{s}}^{\rm{s}}+(1-p){\mathcal L}_{\rm{s}}^{\rm{c}}\in[{\mathcal L}_{\rm{s}}^{\rm{s}},{\mathcal L}_{\rm{s}}^{\rm{c}}]$.
\end{corollary}
\vspace{-5pt}
\vspace{-5pt}
\begin{remark}
The above results suggest that any rate-tuple achieved by the time-sharing strategy yields the same diversity order and high-SNR slope.
\end{remark}
\vspace{-5pt}
\section{Performance of FDSAC}
We consider FDSAC as a baseline scenario, where $\alpha\in[0,1]$ fraction of the total bandwidth is used for communications and the other is used for sensing. On the basis of \cite{Ouyang2022_CL,Ouyang2022_WCL}, the sum-CR and the SR are given by $\overline{\mathcal{R}}_{\rm{c}}^{\rm{f}}=\alpha\log_2\det(\mathbf{I}+{p_{\rm{c}}}{\alpha^{-1}}{{\mathbf{H}}}{{\mathbf{H}}}^{\mathsf{H}})$ and $\mathcal{R}_{\rm{s}}^{\rm{f}}=\max\nolimits_{\mathsf{tr}\left({\mathbf{S}}{\mathbf{S}}^{\mathsf{H}}\right)\leq Lp_{\rm{s}}}\frac{M(1-\alpha)}{L}\log_2\det(\mathbf{I}+\frac{1}{1-\alpha}{\mathbf{S}}^{\mathsf{H}}{\mathbf{R}}
{\mathbf{S}})$, respectively. Accordingly, we derive the following corollary.
\vspace{-5pt}
\begin{corollary}\label{FDSAC_Diversity_Order}
As $p_{\rm c}\rightarrow\infty$, the OP of the sum-CR achieved by FDSAC satisfies $\Pr(\overline{\mathcal{R}}_{\rm{c}}^{\rm{f}}<\mathcal{R})\simeq{\mathcal{O}}(p_{\rm{c}}^{-MK})$.
\end{corollary}
\vspace{-5pt}
\begin{IEEEproof}
Similar to the proof of Theorem \ref{OP_C_SIC_Theorem}.
\end{IEEEproof}
\vspace{-5pt}
\begin{corollary}\label{FDSAC_HSPO}
The high-SNR slopes of $\mathcal{R}_{\rm{c}}^{\rm{f}}={\mathbbmss{E}}\{\overline{\mathcal{R}}_{\rm{c}}^{\rm{f}}\}$ and $\mathcal{R}_{\rm{s}}^{\rm{f}}$ are given by $\alpha K$ and $(1-\alpha)\frac{NM}{L}$, respectively.
\end{corollary}
\vspace{-5pt}
\begin{IEEEproof}
Similar to the proofs of Theorems \ref{Theorem_ECR_C_SIC_Asymptotic} and \ref{Theorem_SR_ER}.
\end{IEEEproof}
Moreover, the rate region achieved by FDSAC is given by
{\setlength\abovedisplayskip{2pt}
\setlength\belowdisplayskip{2pt}
\begin{align}
\mathcal{C}_{\rm{f}}=\left\{\left({\mathcal{R}}^{\rm{s}},{\mathcal{R}}^{\rm{c}}\right)|{\mathcal{R}}^{\rm{s}}\!\in\!\left[0,\!\mathcal{R}_{\rm{s}}^{\rm f}\right],{\mathcal{R}}^{\rm{c}}\!\in\!\left[0,\!\mathcal{R}_{\rm{c}}^{\rm f}\right],\alpha\!\in\!\left[0,\!1\right]\right\}.\label{Rate_Regio_FDSAC}
\end{align}
}After completing all the analyses, we summarize the results related to diversity order and high-SNR slope in Table \ref{table1}.
\begin{table}[!t]
\centering
\setlength{\abovecaptionskip}{0pt}
\resizebox{0.25\textwidth}{!}{
\begin{tabular}{|c|cc|c|}
\hline
\multirow{2}{*}{System} & \multicolumn{2}{c|}{Sum-CR}                        & SR                       \\ \cline{2-4}
                        & \multicolumn{1}{c|}{$\mathcal{D}$} & $\mathcal{S}$ & $\mathcal{S}$            \\ \hline
ISAC (S-SIC)            & \multicolumn{1}{c|}{$MK$}    & $K$           & ${NM}/{L}$           \\ \hline
ISAC (C-SIC)            & \multicolumn{1}{c|}{$MK$}    & $K$           & ${NM}/{L}$           \\ \hline
ISAC (Time-Sharing)     & \multicolumn{1}{c|}{$MK$}    & $K$           & ${NM}/{L}$           \\ \hline
FDSAC                   & \multicolumn{1}{c|}{$MK$}    & $\alpha K$    & $(1-\alpha){NM}/{L}$ \\ \hline
\end{tabular}}
\caption{Diversity Order ($\mathcal{D}$) and High-SNR Slope ($\mathcal{S}$)}
\label{table1}
\vspace{-15pt}
\end{table}
\vspace{-5pt}
\begin{remark}\label{High_SNR_Slope_Compare}
The results in Table \ref{table1} suggest that ISAC and FDSAC yield the same diversity order in terms of the sum-CR. Moreover, since $\alpha\in[0,1]$, we find that ISAC achieves larger high-SNR slopes than FDSAC, which means that ISAC provides more degrees of freedom than FDSAC \cite{Heath2018}.
\end{remark}
\vspace{-5pt}
\section{Numerical Results}
In this section, computer simulation results are used to verify the accuracy of the developed results. The main simulation parameters are set as follows: $M=3$, $N=3$, $K=3$, $L=4$, $\alpha_1=0.1$, $\alpha_2=0.5$, $\alpha_3=1$, and the eigenvalues of $\mathbf{R}$ are given by $\{1,0.1,0.05\}$.

\begin{figure}[!t]
    \centering
    \subfigbottomskip=0pt
	\subfigcapskip=-5pt
\setlength{\abovecaptionskip}{0pt}
    \subfigure[OP for ${\mathcal{R}}=5$ bps/Hz.]
    {
        \includegraphics[height=0.17\textwidth]{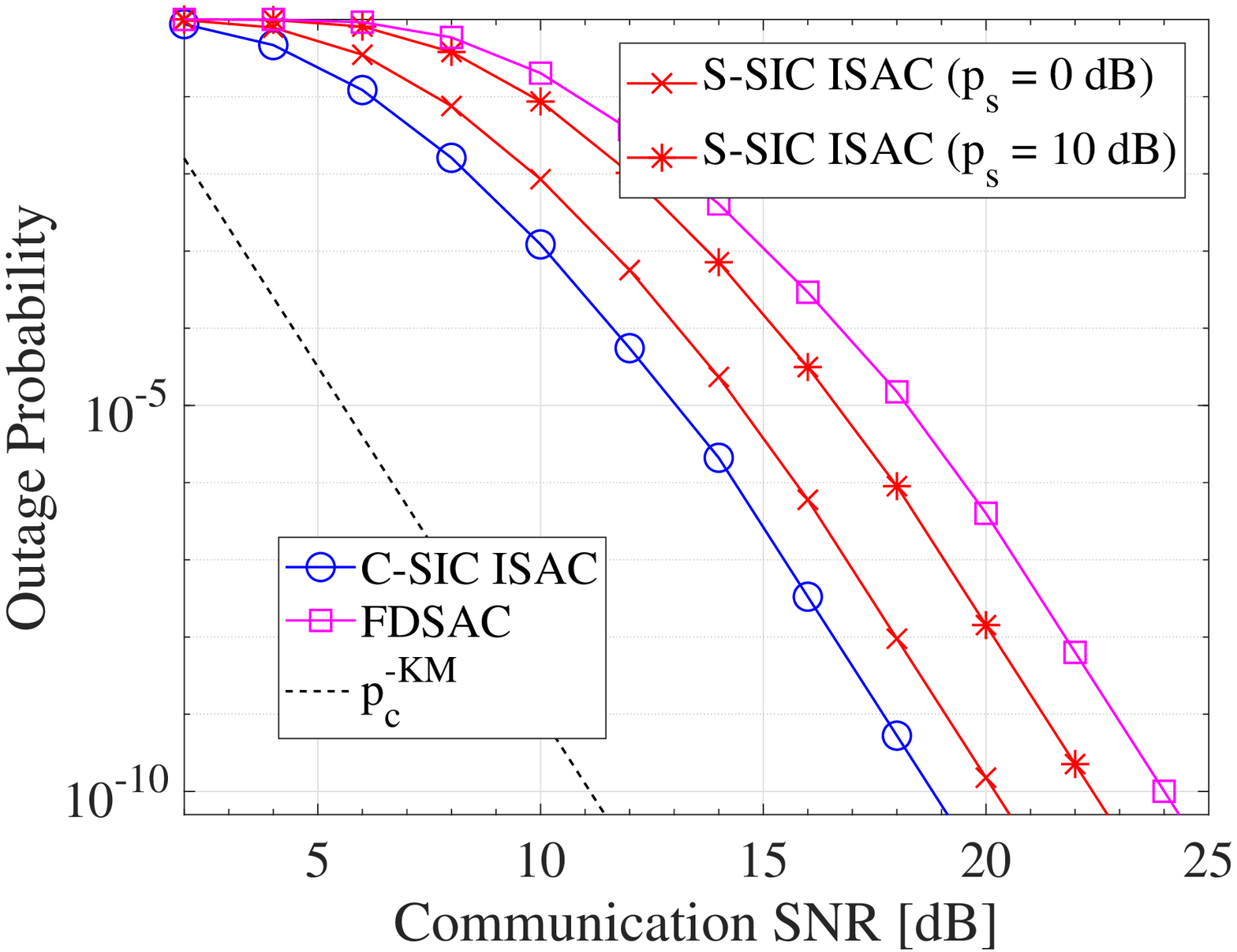}
	   \label{fig2a}	
    }
   \subfigure[Sum ECR.]
    {
        \includegraphics[height=0.17\textwidth]{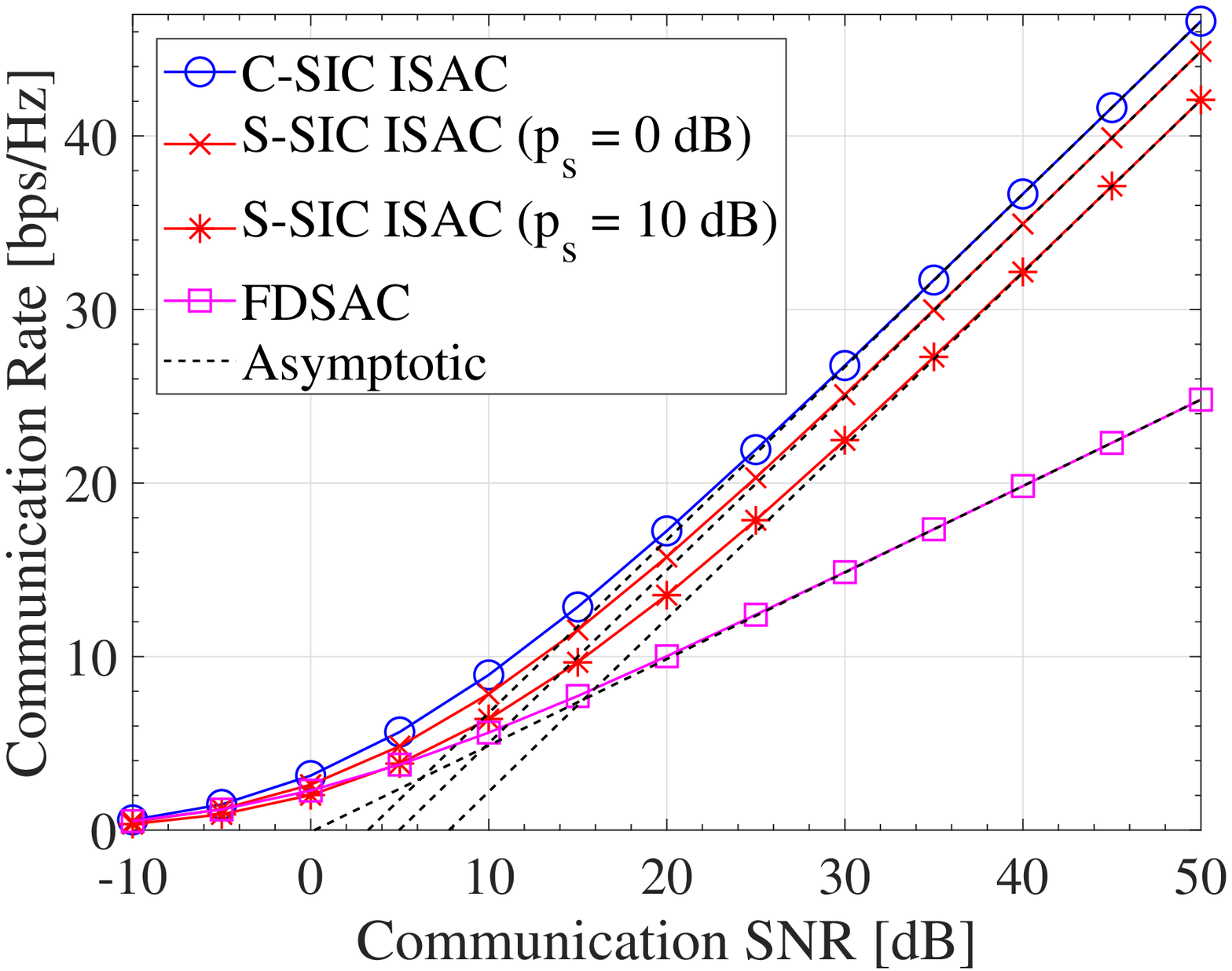}
	   \label{fig2b}	
    }
\caption{Performance of communications. $\alpha=0.5$.}
    \label{Figure2}
    \vspace{-10pt}
\end{figure}

{\figurename} {\ref{fig2a}} plots the OP as a function of the communication SNR $p_{\rm{c}}$. As shown, C-SIC ISAC and FDSAC achieve the lowest and highest OP, respectively. In the high-SNR region, the OP curves for ISAC and FDSAC are parallel to that denoting $p_{\rm{c}}^{-KM}$. This observation means that the presented four cases yield the same diversity order of $KM$. This, together with the fact that C-SIC yields a lower OP than S-SIC, suggests that C-SIC achieves a larger array gain than S-SIC, which supports our discussions in Section \ref{Discussion_SIC_Order}. {\figurename} {\ref{fig2b}} plots the sum ECR versus $p_{\rm{c}}$. It can be seen that the asymptotic results accurately track the provided simulation results in the high-SNR regime. Moreover, C-SIC ISAC and S-SIC ISAC achieve the same high-SNR slope that is larger than that achieved by FDSAC, which, thus, verifies Remark \ref{High_SNR_Slope_Compare}. As {\figurename} {\ref{fig2b}} shows, when achieving the same CR in the high-SNR region, C-SIC is superior to S-SIC by a constant SNR gap, and this gap increases with $p_{\rm{s}}$. This is consistent with our discussions on SIC ordering in Remark \ref{Discussion_SIC_Communications}.

\begin{figure}[!t]
    \centering
    \subfigbottomskip=0pt
	\subfigcapskip=-5pt
\setlength{\abovecaptionskip}{0pt}
    \subfigure[$\alpha=0.5$.]
    {
        \includegraphics[height=0.17\textwidth]{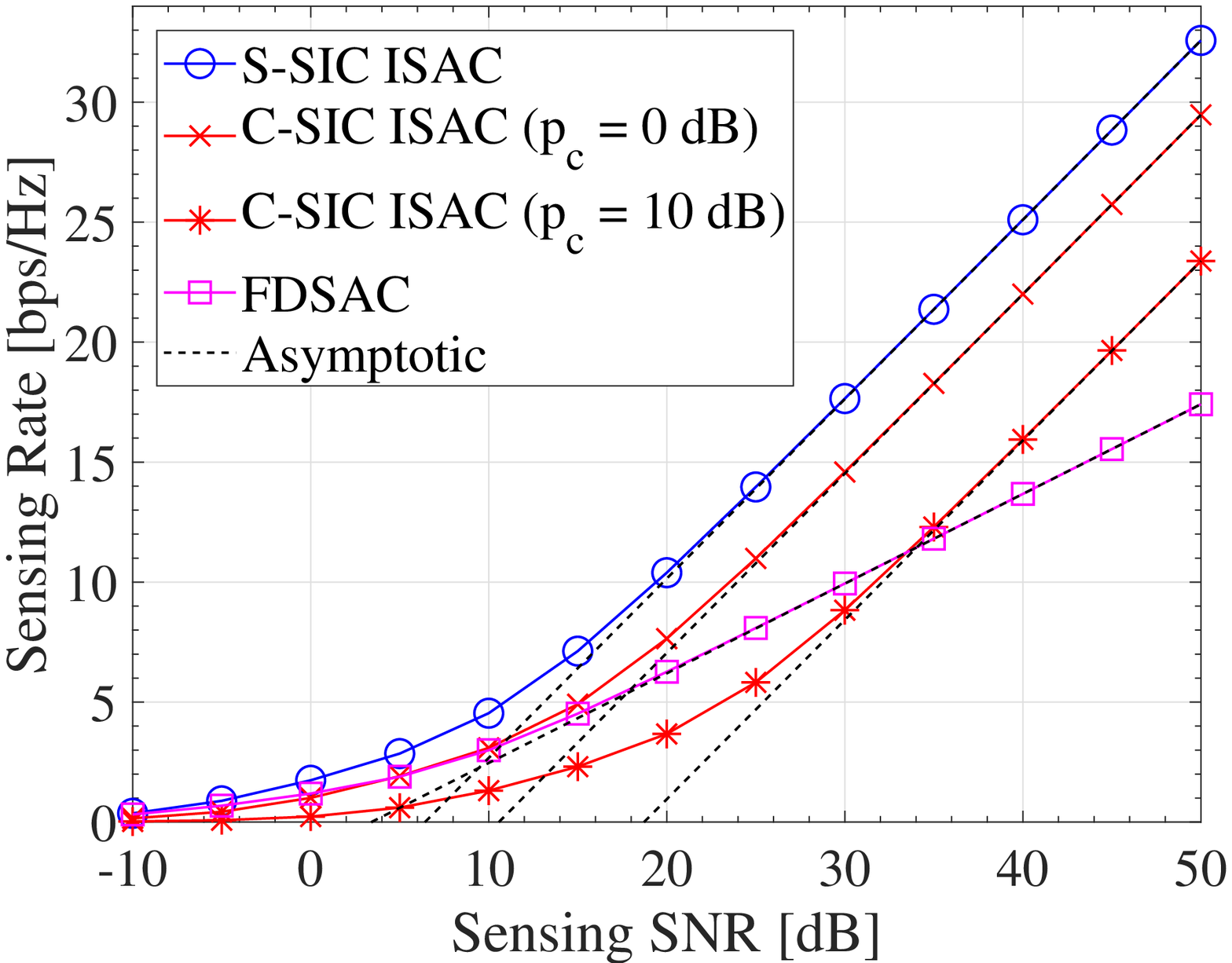}
	   \label{Figure3}	
    }
   \subfigure[$p_{\rm c}=5$ dB and $p_{\rm s}=5$ dB.]
    {
        \includegraphics[height=0.17\textwidth]{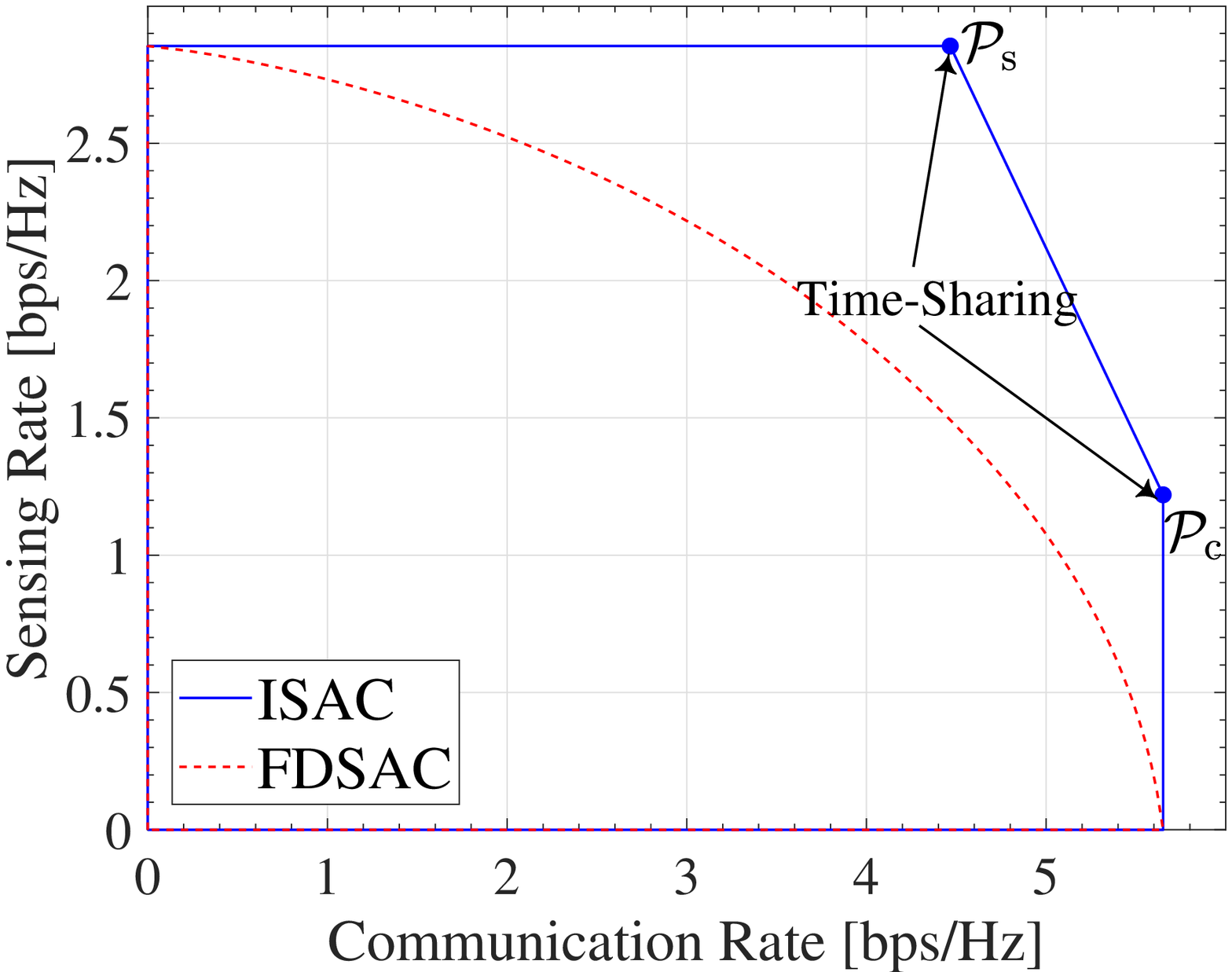}
	   \label{Figure4}	
    }
\caption{(a) Performance of sensing; (b) Rate region.}
    \vspace{-20pt}
\end{figure}

{\figurename} {\ref{Figure3}} plots the SR in terms of the sensing SNR $p_{\rm{s}}$. As expected, S-SIC ISAC achieves the largest SR whereas FDSAC achieves the lowest high-SNR SR. We observe that the asymptotic results match well with the simulation results in the high-SNR regime. Besides, ISAC achieves a larger high-SNR slope than FDSAC. Notably, when achieving the same SR in the high-SNR region, S-SIC is superior to C-SIC by a constant SNR gap, and this gap is more highlighted for a larger value of $p_{\rm{c}}$. This is consistent with our discussions in Remark \ref{Discussion_SIC_Sensing}. {\figurename} {\ref{Figure4}} further compares the S\&C rate regions achieved by FDSAC and ISAC. For ISAC, the points ${\mathcal{P}}_{\rm{s}}$ and ${\mathcal{P}}_{\rm{c}}$ are achieved by the S-SIC and C-SIC, respectively. The line segment connecting ${\mathcal{P}}_{\rm{s}}$ and ${\mathcal{P}}_{\rm{c}}$ is achieved by the time-sharing strategy. As expected, ${\mathcal{P}}_{\rm{s}}$ and ${\mathcal{P}}_{\rm{c}}$ achieve the largest SR and CR, respectively, which reflects the influence of the SIC ordering. Most importantly, we find that the rate region of FDSAC is completely contained within that of ISAC, which highlights the superiority of ISAC.

\section{Conclusion}
In this letter, we have analyzed the uplink S\&C performance achieved by NOMA-ISAC. The diversity orders, high-SNR slopes, and high-SNR power offsets have been derived to discuss the influence of SIC ordering. Analytical and numerical results have shown that SIC order influences SR and CR by shaping the high-SNR power offsets and array gains.
\begin{appendix}
\subsection{Proof of Lemma \ref{Sensing_MI_C_SIC_Lemma}}\label{Proof_Sensing_MI_C_SIC_Lemma}
Let ${\mathbf{z}}_m^{\mathsf H}$ and ${\mathbf{n}}_m^{\mathsf H}$ denote the $m$th row of $\mathbf Z$ and ${\mathbf N}^{\mathsf H}$, respectively. We have ${\mathbf n}_m\sim{\mathcal{CN}}({\mathbf 0},{\mathbf I})$ and ${\mathbf{z}}_m=\sum_{k=1}^{K}h_{k,m}^{*}{\mathbf x}_{k}+{\mathbf{n}}_m$. It follows that ${\mathbbmss E}\{{\mathbf{z}}_m\}=0$ and ${\mathbbmss E}\{{\mathbf{z}}_m{\mathbf{z}}_m^{\mathsf H}\}={\mathbf I}+\sum_{k=1}^{K}{\mathbbmss E}\{\lvert h_{k,m}\rvert^2\}{\mathbbmss E}\{{\mathbf x}_{k}{\mathbf x}_k^{\mathsf H}\}=(1+\sum_{k=1}^{K}\alpha_k{p_{\rm{c}}}){\mathbf I}=\sigma_{\rm{c}}^2{\mathbf I}$. Moreover, for $m\neq m'$, we have ${\mathbbmss E}\{{\mathbf{z}}_m{\mathbf{z}}_{m'}^{\mathsf H}\}={\mathbbmss E}\{{\mathbf{n}}_m{\mathbf{n}}_{m'}^{\mathsf H}\}+\sum_{k=1}^{K}{\mathbbmss E}\{ h_{k,m}^{*}h_{k,m'}\}{\mathbbmss E}\{{\mathbf x}_{k}{\mathbf x}_k^{\mathsf H}\}={\mathbf 0}$. Thus, when $\mathbf Z$ is treated as Gaussian noise, it contains $LM$ i.i.d. CGEs each with zero mean and variance $\sigma_{\rm{c}}^2$. Using similar steps as those outlined in \cite[Appendix C]{Ouyang2022_WCL}, we can obtain Lemma \ref{Sensing_MI_C_SIC_Lemma}.
\vspace{-10pt}
\subsection{Proof of Theorem \ref{Theorem_ECR_C_SIC_Asymptotic}}\label{Proof_Theorem_ECR_C_SIC_Asymptotic}
As $p_{\rm{c}}\rightarrow\infty$, we have ${\mathcal{R}}_{{\rm{c}}}^{\rm{c}}\approx{\mathbbmss E}\{\log_2\det(p_{\rm{c}}{{\mathbf{H}}}^{\mathsf{H}}{{\mathbf{H}}})\}$. Note that ${\mathbf{H}}=\overline{\mathbf{H}}\mathsf{diag}\{\sqrt{\alpha_1},\ldots,\sqrt{\alpha_K}\}$ with $\overline{\mathbf{H}}$ containing $KM$ i.i.d. standard CGEs. It follows that ${\mathbbmss E}\{\log_2\det(p_{\rm{c}}{{\mathbf{H}}}^{\mathsf{H}}{{\mathbf{H}}})\}={\mathbbmss E}\{\log_2\det(\overline{{\mathbf{H}}}^{\mathsf{H}}\overline{{\mathbf{H}}})\}+K\log_2{p_{\rm c}}+\sum_{k=1}^{K}\log_2{\alpha_k}$. With the aid of \cite[Eq. (C.28)]{Heath2018}, the final results can be obtained.
\vspace{-10pt}
\subsection{Proof of Lemma \ref{Commun_MI_S_SIC_Lemma}}\label{Proof_Commun_MI_S_SIC_Lemma}
Note that ${\mathbbmss{E}}\{{\mathbf a}_l\}={\mathbbmss E}\{{\mathbf{G}}^{\mathsf{H}}{\mathbf{s}}_l\}+{\mathbbmss E}\{{\mathbf{n}}_{l}\}=0$ and ${\mathbbmss E}\{{\mathbf a}_l{\mathbf a}_l^{\mathsf H}\}={\mathbbmss E}\{{\mathbf{G}}^{\mathsf{H}}{\mathbf{s}}_l{\mathbf{s}}_l^{\mathsf H}{\mathbf G}\}+{\mathbf I}$. Based on the statistics of $\mathbf{G}=[\mathbf{g}_{1} \ldots \mathbf{g}_M]$, we have ${\mathbf g}_m^{\mathsf H}{\mathbf s}_l\sim{\mathcal {CN}}(0,\lvert{\mathbf{s}}_l^{\mathsf{H}}{\mathbf{R}}{\mathbf{s}}_l\rvert)$. Since ${\mathbf{g}}_m$ is independent with ${\mathbf g}_{m'}$ for $m'\neq m$, we have ${\mathbbmss E}\{{\mathbf{G}}^{\mathsf{H}}{\mathbf{s}}_l{\mathbf{s}}_l^{\mathsf H}{\mathbf G}\}=\lvert{\mathbf{s}}_l^{\mathsf{H}}{\mathbf{R}}{\mathbf{s}}_l\rvert{\mathbf I}$. Thus, when ${\mathbf a}_l$ is treated as Gaussian noise, it contains $M$ i.i.d. CGEs each with zero mean and variance $\varrho_l^2=1+|{\mathbf{s}}_l^{\mathsf{H}}{\mathbf{R}}{\mathbf{s}}_l|$. The final results follow immediately.
\vspace{-10pt}
\end{appendix}

\end{document}